\documentclass[11pt,twoside]{article}
\usepackage{asp2010}
\usepackage{natbib}
\usepackage{hyperref}
\resetcounters

\begin{document}

\allowtitlefootnote

\title{On Nuclear Matter Cores and Their Applications\footnotemark}
\author{~K.~Boshkayev,$^{1,~2}$, ~M.~Rotondo,$^{1,~2}$ and R.~Ruffini$^{1,~2}$
\affil{$^1$Physics Department, University of Rome "Sapienza", Square Aldo Moro 5, I-00185 Rome, Italy\\
$^2$ICRANet, Square of the Republic 10, I-65122 Pescara, Italy\\
}}


\begin{abstract}
We review recent series of articles considering electromagnetic effects in self-gravitating systems of nuclear matter. The results find their explicit application within the theory of neutron stars. 
\end{abstract}

The Feynman-Metropolis-Teller treatment of compressed atoms has been recently generalized to the relativistic regimes by ~\cite{2011PhRvC..83d5805R}. The condition of $\beta$-equilibrium between neutrons, protons, and electrons has been there enforced self-consistently in the relativistic Thomas-Fermi equation extending also the works of \cite{popov71}, \cite{zeldovich72}, \cite{migdal76, migdal77}, \cite{1980PhLB...91..314F} and \cite{1981PhLB..102..442R} for heavy nuclei. Thanks to the existence of scaling laws (see \cite{2008pint.conf..207R}) this treatment has been extrapolated to nuclear matter cores of stellar dimensions, i.e. degenerate systems consisting of $N_n$ neutrons, $N_p$ protons and $N_e$ electrons in $\beta$-equilibrium at nuclear densities having mass numbers $A=N_p+N_n \simeq (m_{\rm Planck}/m_n)^3 \sim 10^{57}$ and correspondingly $M_{core}\sim M_{\odot}$. The case in absence of external forces was first addressed by \cite{Ruffini2007IJMPD..16....1R} and \cite{2011IJMPD..20.1995R} while, the compressed case, was introduced in \cite{2011PhRvC..83d5805R}. Such configurations fulfill global, but not local charge neutrality, i.e. $N_e=N_p$ and have electric fields on the core surface larger than the critical value $E_c = m_e^2c^3/(e\hbar)$. A constant distribution of protons at nuclear densities has been assumed which simulates, in such a treatment, the confinement due to the strong interactions in the case of nuclei and heavy nuclei and due to both the gravitational field and the strong interactions in the case of nuclear matter cores of stellar dimensions.

The explicit treatment of the effects of gravity on these systems has been recently considered in \cite{rotondoPLB2011}. The self-consistent treatment of the simplest, nontrivial, self-gravitating system of degenerate neutrons, protons and electrons in $\beta$-equilibrium within relativistic quantum statistics and the Einstein-Maxwell equations has been there presented. The impossibility of imposing the condition of local charge neutrality on such systems has been proved, consequently overcoming the Tolman-Oppenheimer-Volkoff treatment traditionally assumed in the neutron stars theory. The crucial role of the constancy of the generalized Fermi energies for each particle species within the system has been demonstrated in complete generality. A new approach based on the coupled system of the general relativistic Thomas-Fermi equations and the Einstein-Maxwell equations has been presented and an explicit solution fulfilling global but not local charge neutrality has been obtained.

The extension of the equilibrium equations presented in \cite{rotondoPLB2011} to the case of finite temperatures has been illustrated in \cite{RRRX2011}. The constancy of the generalized chemical potentials of particles, briefly the Klein potentials, in the case of finite temperatures generalizing the condition of the general relativistic Fermi energies in the case of zero temperature has been there demonstrated. The further crucial step of taking into due account the strong interactions between nucleons has been also recently achieved \citep{2011NuPhA.872..286R}.

What about rotation? In reality neutron stars are observed as a consequence of the effects emerging due to their rotation. The first attempt to take into consideration the rotation of the nuclear matter cores of stellar dimensions was made in \cite{2012IJMPS..12...58B} in the frame of classical electrodynamics, using the technique developed by \cite{Marsh1982AmJPh..50...51M}. There we have considered the magnetic field induced by rigid rotation, generalizing in this way a part of the work of \cite{2011IJMPD..20.1995R}. According to \cite{2012IJMPS..12...58B} the magnetic field evolving due to rigid rotation has some special peculiarities like the electric field of the system although the system is globally neutral. Firstly, near the surface of the cores the magnitude of the radial component of the magnetic field $B_r$ is much smaller than the tangential component of the magnetic field $B_{\theta}$ whose value is overcritical at the equatorial plane. Secondly, outside the cores these quantities fall down very quickly. Inside the cores their magnitudes are constant and equal to each other and the value of them are very small with respect to the critical magnetic field $B_c=4.4.\times 10^{13}$ G. Thirdly, like the electric field, the magnetic field is mainly concentrated in a very thin shell on the surface of the core. Following the work of  \cite{2012IJMPS..12...58B} the stability of rotating nuclear matter cores against the rotational kinetic energy and induced magnetic energy was investigated in \cite{2010tsra.confE.275B}. In fact the whole system turned out to be gravitationally bound and stable even for millisecond rotational periods. The magnetic energy of the thin shell has the order of one tenth of the rotational energy of the same shell irrespective of the value of the rotational period of the core.

All the above results can be applied to the physics of compact objects; e.g. the emergence of critical electromagnetic fields in neutron stars and the process of gravitational collapse to a black hole. We are currently considering the effects of magnetic fields and the stability of rotating systems of nuclear matter in general relativity, generalizing the results in \cite{rotondoPLB2011, RRRX2011, 2011NuPhA.872..286R}. 


\begin{thebibliography}{}
\expandafter\ifx\csname natexlab\endcsname\relax\def\natexlab#1{#1}\fi
\expandafter\ifx\csname url\endcsname\relax
  \def\url#1{\texttt{#1}}\fi
\expandafter\ifx\csname urlprefix\endcsname\relax\def\urlprefix{URL }\fi
\providecommand{\eprint}[2][]{\url{#2}}

\bibitem[{{Boshkayev} et~al.(2010){Boshkayev}, {Rotondo}, \&
  {Ruffini}}]{2010tsra.confE.275B}
{Boshkayev}, K., {Rotondo}, M., \& {Ruffini}, R. 2010, in 25th Texas Symposium
  on Relativistic Astrophysics

\bibitem[{{Boshkayev} et~al.(2012){Boshkayev}, {Rotondo}, \&
  {Ruffini}}]{2012IJMPS..12...58B}
--- 2012, International Journal of Modern Physics Conference Series, 12, 58

\bibitem[{{Ferreirinho} et~al.(1980){Ferreirinho}, {Ruffini}, \&
  {Stella}}]{1980PhLB...91..314F}
{Ferreirinho}, J., {Ruffini}, R., \& {Stella}, L. 1980, Physics Letters B, 91,
  314

\bibitem[{{Marsh}(1982)}]{Marsh1982AmJPh..50...51M}
{Marsh}, J.~S. 1982, American Journal of Physics, 50, 51

\bibitem[{{Migdal} et~al.(1977){Migdal}, {Popov}, \& {Voskresenski{\v
  i}}}]{migdal77}
{Migdal}, A.~B., {Popov}, V.~S., \& {Voskresenski{\v i}}, D.~N. 1977, Soviet
  Journal of Experimental and Theoretical Physics, 45, 436

\bibitem[{{Migdal} et~al.(1976){Migdal}, {Voskresenski{\v i}}, \&
  {Popov}}]{migdal76}
{Migdal}, A.~B., {Voskresenski{\v i}}, D.~N., \& {Popov}, V.~S. 1976, Soviet
  Journal of Experimental and Theoretical Physics Letters, 24, 163

\bibitem[{{Popov}(1971)}]{popov71}
{Popov}, V.~S. 1971, Soviet Journal of Experimental and Theoretical Physics,
  32, 526

\bibitem[{{Rotondo} et~al.(2011{\natexlab{a}}){Rotondo}, {Rueda}, {Ruffini}, \&
  {Xue}}]{RRRX2011}
{Rotondo}, M., {Rueda}, J.~A., {Ruffini}, R., \& {Xue}, S.-S.
  2011{\natexlab{a}}, ArXiv e-prints. \eprint{1107.2777}

\bibitem[{{Rotondo} et~al.(2011{\natexlab{b}}){Rotondo}, {Rueda}, {Ruffini}, \&
  {Xue}}]{2011PhRvC..83d5805R}
--- 2011{\natexlab{b}}, \prc, 83, 045805. \eprint{0911.4622}

\bibitem[{{Rotondo} et~al.(2011{\natexlab{c}}){Rotondo}, {Rueda}, {Ruffini}, \&
  {Xue}}]{rotondoPLB2011}
--- 2011{\natexlab{c}}, Physics Letters B, 701, 667. \eprint{1106.4911}

\bibitem[{{Rotondo} et~al.(2011{\natexlab{d}}){Rotondo}, {Ruffini}, {Xue}, \&
  {Popov}}]{2011IJMPD..20.1995R}
{Rotondo}, M., {Ruffini}, R., {Xue}, S.-S., \& {Popov}, V. 2011{\natexlab{d}},
  International Journal of Modern Physics D, 20, 1995

\bibitem[{{Rueda} et~al.(2011){Rueda}, {Ruffini}, \&
  {Xue}}]{2011NuPhA.872..286R}
{Rueda}, J.~A., {Ruffini}, R., \& {Xue}, S.-S. 2011, Nuclear Physics A, 872,
  286. \eprint{1104.4062}

\bibitem[{{Ruffini}(2008)}]{2008pint.conf..207R}
{Ruffini}, R. 2008, in Path Integrals - New Trends and Perspectives, 207

\bibitem[{{Ruffini} et~al.(2007){Ruffini}, {Rotondo}, \&
  {Xue}}]{Ruffini2007IJMPD..16....1R}
{Ruffini}, R., {Rotondo}, M., \& {Xue}, S.-S. 2007, International Journal of
  Modern Physics D, 16, 1. \eprint{arXiv:astro-ph/0609190}

\bibitem[{{Ruffini} \& {Stella}(1981)}]{1981PhLB..102..442R}
{Ruffini}, R., \& {Stella}, L. 1981, Physics Letters B, 102, 442

\bibitem[{{Zeldovich} \& {Popov}(1972)}]{zeldovich72}
{Zeldovich}, Y.~B., \& {Popov}, V.~S. 1972, Soviet Physics Uspekhi, 14, 673

\end{thebibliography}

\end{document}